\documentclass{jps-cp}
\usepackage{footmisc}
\usepackage{graphicx}
\usepackage{siunitx}
\usepackage{booktabs}
\usepackage{geometry}
\newcommand{\exgra}{e.g..\ }

\title{Systematic Approach for Tuning Flux-driven Josephson Parametric
Amplifiers for Stochastic Small Signals}

\author{
    \c{C}a\u{g}lar \textsc{Kutlu}$^{1,2}$,
    Saebyeok \textsc{Ahn}$^{1,2}$,
    Sergey V. \textsc{Uchaikin}$^{2}$,
    Soohyung \textsc{Lee}$^{2}$,
    Arjan F. \textsc{van Loo}$^{3,4}$,
    Yasunobu \textsc{Nakamura}$^{3,4}$,
    Seonjeong \textsc{Oh}$^{2}$,
    and Yannis K. \textsc{Semertzidis}$^{2,1}$
}

\inst{
    $^{1}$Korea Advanced Institute of Science and Technology, Daejeon 34051, Republic of Korea\\
    $^{2}$Center for Axion and Precision Physics Research, Institute for Basic Science, Daejeon, 34051, Republic of Korea \\
    $^{3}$RIKEN Center for Quantum Computing (RQC), Wako, Saitama 351–0198, Japan \\
    $^{4}$Department of Applied Physics, Graduate School of Engineering, The University of Tokyo, Bunkyo-ku, Tokyo 113-8656, Japan
}

\email{caglar.kutlu@gmail.com; uchaikin@ibs.re.kr}

\recdate{August 2, 2022}

\abst{Many experiments operating at millikelvin temperatures with signal frequencies
in the microwave regime are beginning to incorporate Josephson Parametric
Amplifiers (JPA) as their first amplification stage. While there are
implementations for a wideband frequency response with a minimal need for tuning,
designs using resonant structures with small numbers of Josephson elements
still achieve the best noise performance.  In a typical measurement scheme
involving a JPA, one needs to control the resonance frequency, pump frequency and
pump power to achieve the desired amplification and noise properties.  In this
work, we propose a straightforward approach for operating JPAs with the help of
a look-up table (LUT) and online fine-tuning.  Using the proposed approach, we
demonstrate the operation of a flux-driven JPA with 20 dB gain around 5.9 GHz,
covering approximately 100 MHz with 20 kHz tuning steps.  The proposed
methodology was successfully used in the context of a haloscope axion
experiment.}

\kword{josephson parametric amplifier, axion dark matter, noise temperature,
cryogenic microwave amplifier}

\begin{document}
\maketitle

\section{Introduction}
Quantum mechanics imposes a minimum uncertainty constraint in all measurement systems.
This property manifests itself in the context of microwave detection chains as the
quantum limit on added noise and given by $ T_Q = \frac{hf}{2k_B} $ \cite{Caves1982}.
While there have been demonstrations of techniques evading this limit\cite{Zhong2013}
(\exgra squeezing), they typically require the phase of the expected signal to be
known\cite{Clerk2010}.  Therefore, such techniques are not directly applicable to
measurements of signals with a stochastic phase component.  Experiments looking for
signatures from axion-like particles as components for the unknown matter in our galaxy
involve such measurements\cite{BrubakerAxion2017,Lee2020,Jeong2020,Kwon2021}.

By convention, the noise characteristics of detection chains are quantified using their
so-called noise temperature $T_n$, which is defined using the measured power within a
bandwidth $ B $ given as $ P = G(P_\mathrm{src} + k_B B T_{n}) $, where $ G$ is the
power gain of the whole chain, $P_{\mathrm{src}} $ is the source signal with its
accompanying noise and $k_B$ is Boltzmann constant.  Currently, the lowest noise
temperatures in the microwave regime are achieved using Josephson parametric amplifiers
(JPA) as the first stage in the detection chain.  Since the noise temperature of a JPA
depends on its input noise\cite{Kutlu2021}, such systems typically operate in the
millikelvin regime to reduce the thermal background. While there are many competing JPA
designs, this work focuses on a flux-driven JPA design\cite{Yamamoto2008} comprising a $
\lambda/4 $ resonator terminated with a superconducting quantum interference device
(SQUID).  The DC component of the magnetic flux through the SQUID loop controls the
resonance frequency ($ f_r $) while the AC component provides the inductance modulation
necessary to achieve parametric amplification.  The design incorporates a second
transmission line that inductively couples to the SQUID loop.  This path, referred to as
the pump line, provides the AC component of the magnetic flux. An external coil
provides the static magnetic field in the experimental fixture.  In order to operate the JPA 
with the desired amplification and noise properties one needs to tune the coil current ($ i_b
$), pump frequency ($ f_p $), and pump power ($ P_p $).  This work demonstrates a method
for controlling these parameters to achieve the minimum noise temperature for a particular
gain requirement.  The proposed method is implemented in an experiment searching for a
hypothetical particle, the axion, expected to comprise the unknown matter
content in the milky way galaxy.

\section{Methodology}
When the JPA is operated in the three-wave mixing mode, the relation $ f_p = f_s + f_i $ will
be satisfied where $ f_s $ is the signal frequency and the $ f_i $ is the idler
frequency\cite{Roy2016}.
Throughout this work, the output is always measured at the signal frequency.  The JPA
gain is a function of frequency typically with a peak occuring at $ f_{sc} = f_p/2 $.
Using $ G_J $ to denote the peak gain value, the tuning is done by following
the steps below~:

\begin{enumerate}
    \item Tune $ i_b $ such that $f_r$ is equal to the
        desired $ f_{sc} $.
    \item Set $ f_p $ to $2 f_{sc} $.
    \item Increase $ P_p $ until the desired gain is
        obtained.
    \item Repeat step 2 and 3 for \emph{small} deviations from  $ f_r $ by changing $ i_b $.
	\item Pick a set of $ i_b, f_p, P_p $ that has the lowest $ P_p $ for the desired gain at $ f_{sc} $.
\end{enumerate}

While the optimization protocol described above is straightforward, it requires
$ f_{sc} $ to be known ahead of time.  Since for many experiments $ f_{sc}$ is
decided during the experiment, this is not very practical. In order to avoid
spending time on optimization during the experiment, we generate a
look-up table (LUT) for JPA state parameters using $ f_{sc} $ and $G_J$ as
indexing pairs.  Introducing the detuning variable as $\delta = f_p/2 -
f_r$\footnote{Experimentally, we vary $ \delta $ only by tuning $f_p$.}, we
first do a set of characterization measurements with the following steps~:

\begin{enumerate}
    \item With the pump off, measure $f_r$ as a function of $ i_b $.
    \item Given the frequency range of interest [$f_{\mathrm{min}}$,
        $f_{\mathrm{max}}$], measure $ G_J$ at $i_b(f_r=f_{\mathrm{min}})$ and
        $i_b(f_r=f_{\mathrm{max}})$ as a function of $P_p$ and $\delta$.  
        From these measurements, define a rectangular sweeping region for $ \delta $ and
        $ P_p $ where $ G_J \geq \SI{0}{\decibel} $.  
    \item Perform a detailed $G_J$ measurement by sweeping $ \delta $ and $P_p $ at a
        set of $i_b$ covering [$f_{\mathrm{min}}$, $f_{\mathrm{max}}$].  This step
        yields the complete $G_J$ dataset for $ (i_b, f_p, P_p) \rightarrow (f_{sc}, G_J)$.
\end{enumerate}

\begin{figure}[tbh]
    \centering
    \includegraphics[width=0.95\textwidth]{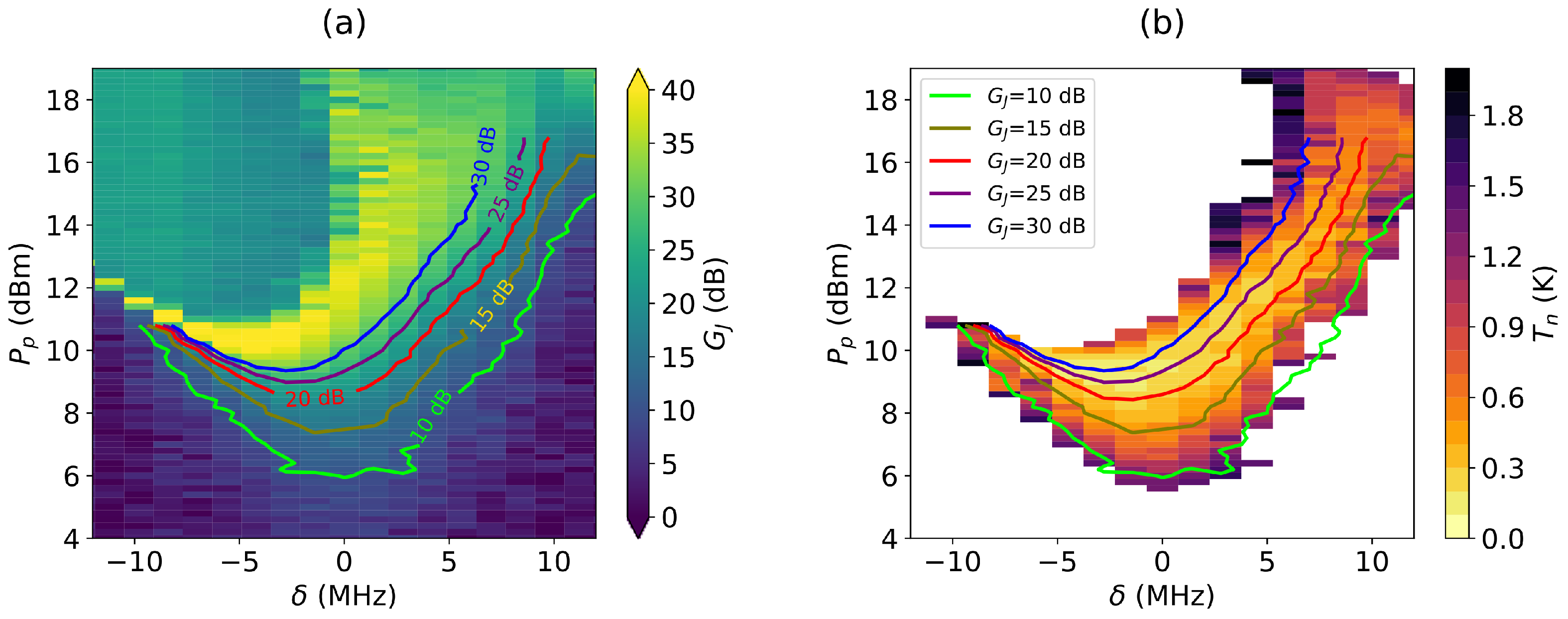}
    \caption{Measurements performed on a test bench setup.  (a) $ G_J$ as a
        function of $ \delta $ and $ P_p $.  The overlaid lines correspond to contours
        calculated from the same data with the upper sections avoided for clarity.  (b)
        $ T_n $ as a function of $ \delta $ and $ P_p $.  The gain contours
        shown in (a) are overlaid for comparison.  It can be seen that noise
        temperatures along a particular equigain curve are minimal at the lowest $ P_p$.}
        \label{fig:ntminimal}
\end{figure}

In order to investigate the noise behavior with respect to $i_b$, $f_p$,
$P_p$, and $G_J$  we have measured $T_n$ as a function of $ f_p$ and $P_p$ for a small
set of $i_b$\footnote{We limit these measurements to a few $i_b$ because it is much more
time consuming to measure $T_n$ in comparison to $G_J$.}.  These measurements reveal
that for a particular $ G_J$, $ T_n$ is minimized when $P_p$ is minimized (see
Fig.~\ref{fig:ntminimal}).  Using this as a constraint we construct the LUT using the
following algorithm~:

\begin{enumerate}
    \item Upsample the $ G_J$ dataset in $ \delta $, $P_p$ and $ i_b $ using linear interpolation.
    \item Using $ f_{sc} = f_p/2 $, group data points in intervals with lengths $ f_{step} $
        in the range [$f_{\mathrm{min}}$, $f_{\mathrm{max}}$].
    \item For each frequency interval, group measurements by their gains
        with the bin edges given by the set $ S_G = \{ G_{1}, G_{2},\dots, G_{N} \}$.
    \item Construct the LUT by picking the $\{i_b, f_p, P_p\}$ with the smallest $ P_p $
        for each gain group within each frequency interval.
    \item If the table is not densely populated enough, repeat from step 1 with higher
        upsampling factors.
\end{enumerate}

The end product of this process is a LUT providing a mapping of the form 
$(f_{sc}, G_J) \rightarrow (i_b, f_{p}, P_{p})$\footnote{One
should make the distinction that the indexes in LUT merely label the intervals.}. 
With access to this table, an index look-up operation is sufficient to find
the noise-optimal operation point for a desired $G_J$.  
The benefit to this approach, as opposed to selecting the desired working point
from the full $G_J$ dataset during experiment is the reduced computational
complexity.

\begin{table}[tbh]
\centering
\includegraphics{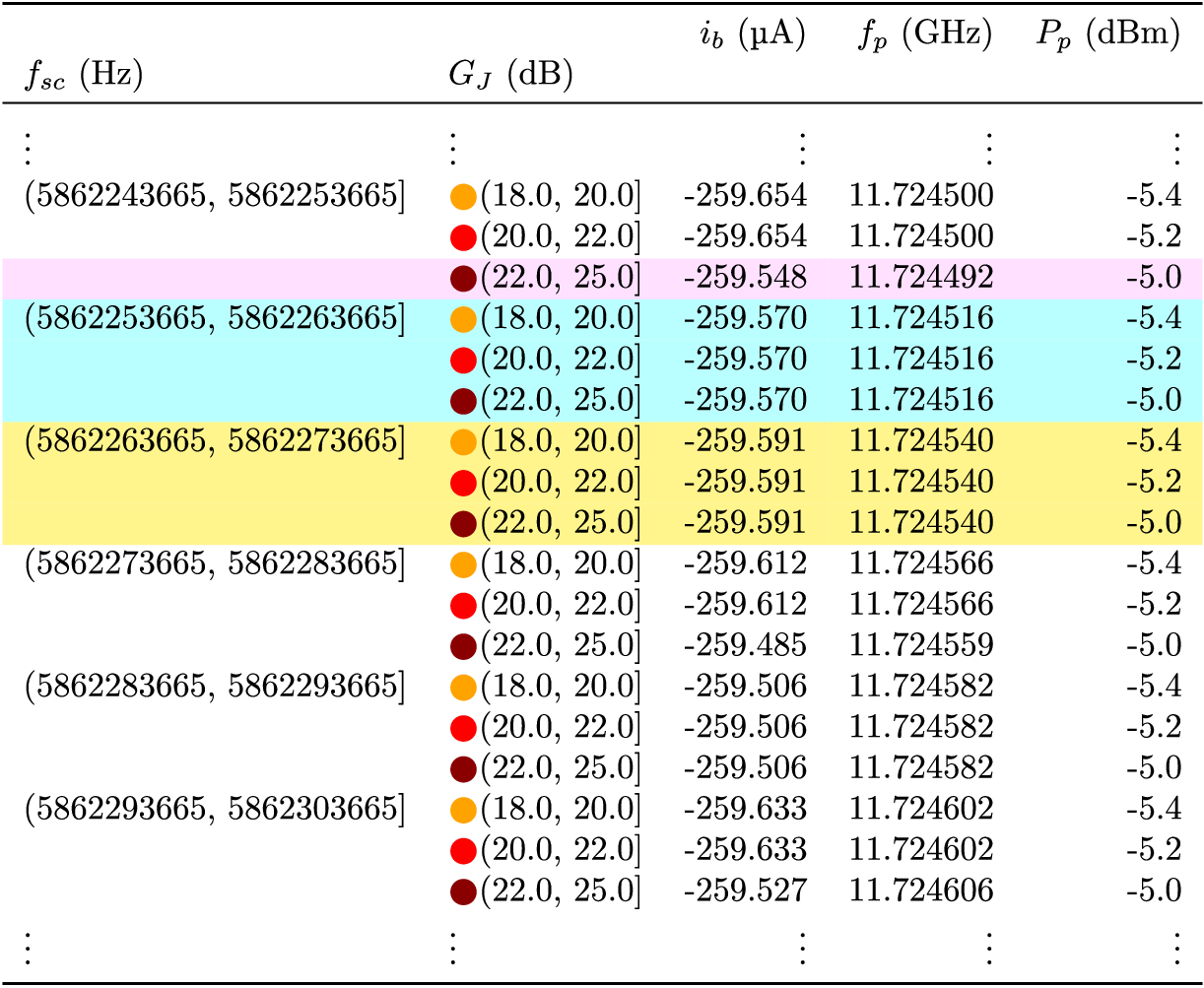}
\caption{A small section from a LUT constructed with the method described in this work.
The colored rows correspond to points contributed from measurements shown in
Fig.~\ref{fig:paramaplut}.}
\label{tab:lutexample}
\end{table}

\begin{figure}[tbh]
    \centering
    \includegraphics[width=0.95\textwidth]{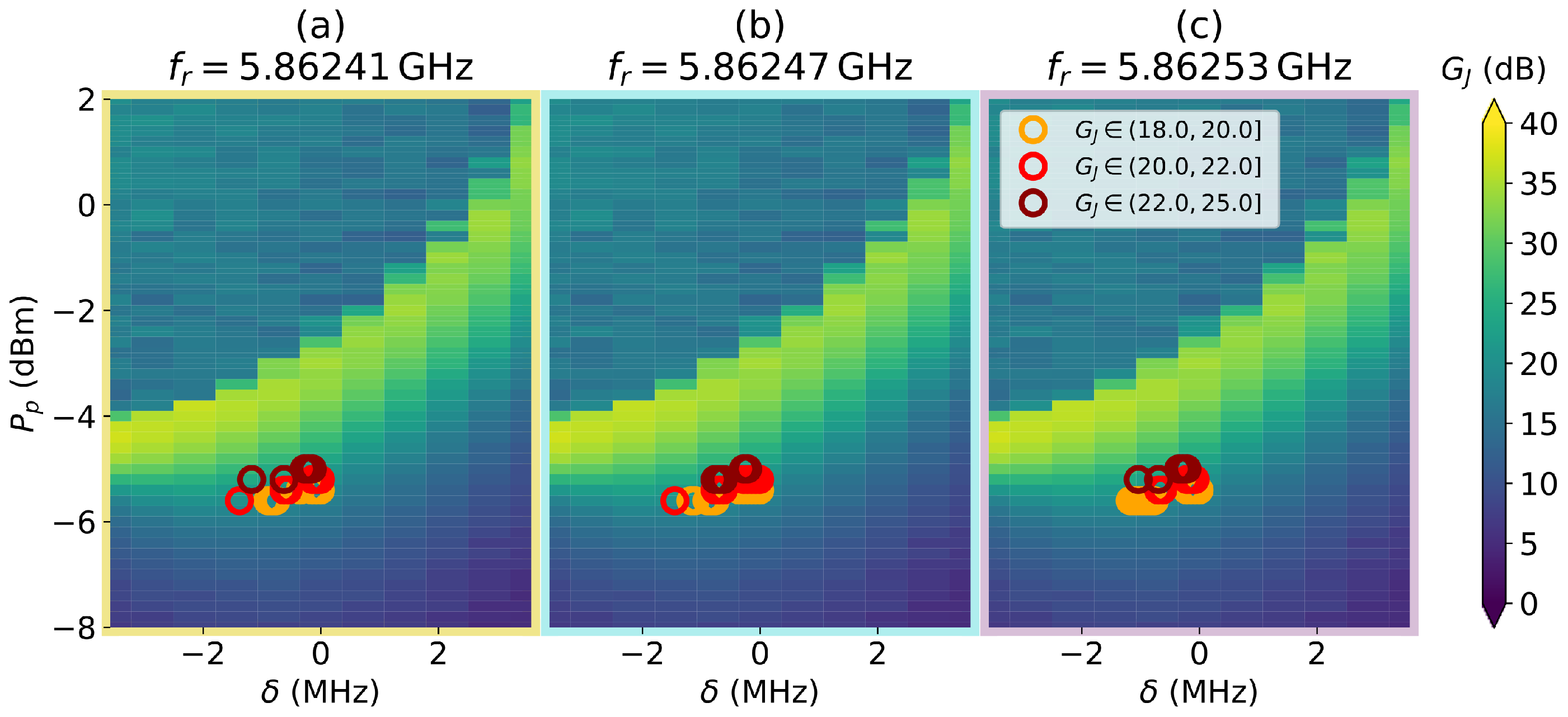}
\caption{$G_J$ as a function of $\delta= f_p/2 - f_r$ and $P_p$ performed at
three example $i_b$ corresponding to $ f_r = \SI{5.86241}{\giga\hertz}$ (a), $
f_r = \SI{5.86247}{\giga\hertz}$ (b) and $ f_r = \SI{5.86253}{\giga\hertz}$ (c).  In a
typical dataset there are a total of 341 such measurements.  The circles correspond to
the points chosen by the algorithm for inclusion in the LUT.}
\label{fig:paramaplut}
\end{figure}

\section{Application}
The described procedure is implemented as part of an axion haloscope in the
Center for Axion and Precision Physics Research (CAPP).
The haloscope consists of a microwave cavity, an \SI{8}{\tesla} superconducting
magnet surrounding it, and a receiver chain to transfer the signal into the
spectrum analyzer.  The experiment is housed in a Bluefors LD400 dilution
refrigerator with the cavity and the JPA installed at the mixing-chamber (MC)
plate.  The MC temperature was kept at \SI{40}{\milli\kelvin} during all of the
experiments.

\begin{figure}[tbh]
    \centering
    \includegraphics[width=0.95\textwidth]{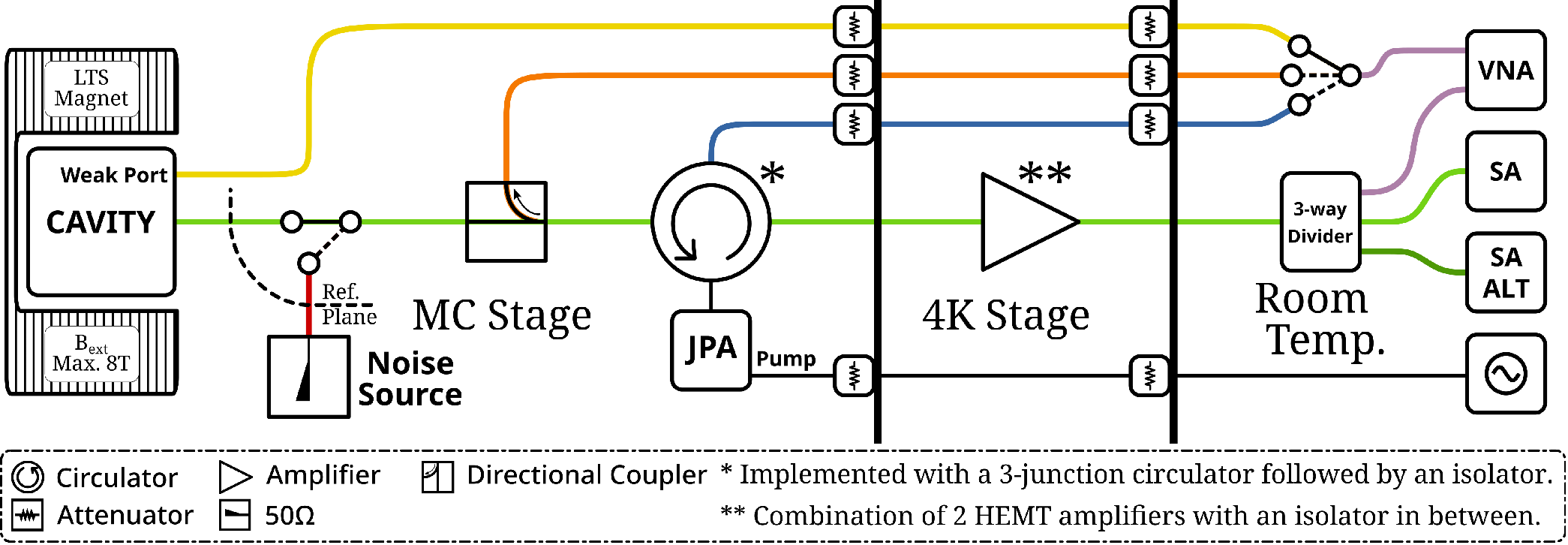}
\caption{Simplified diagram for an axion haloscope.  The colors are used to
highlight several experimentally relevant signal paths.  The JPA characterization
measurements are performed with the room temperature switch pointing at the path shown
in orange.}
\label{fig:expscheme}
\end{figure}

The $G_J$ is estimated by measuring $ s_{21} $ at a single frequency $ f = f_{sc} +
\SI{1}{\kilo\hertz} $ using a vector network analyzer (see Fig.~\ref{fig:expscheme}).
First, a baseline measurement ($s_{\mathrm{off}}$) is performed after tuning $ f_r $
away from the frequency region of interest\footnote{ $ i_b $ is adjusted so that $ \Phi
= -0.48 \Phi_0$. This ensures $ f_r $ is far from the frequency of interest which
typically have $ \Phi > -0.4 $.}.  Then the JPA is tuned to the desired working point,
and a subsequent s-parameter measurement ($s_{\mathrm{off}}$) is performed.  From these
two measurements, the power gain\footnote{One should keep in mind that this is actually
an estimation for the \emph{total gain change} rather than the gain of the JPA itself.}
is estimated via $G_J = 10\log_{10} \left|{s_{\mathrm{on}}}/{s_{\mathrm{off}}}\right|^2
$. Following the procedure described in the previous section, the swept data set for $
G_J $ is obtained for $ f_r $ in the range
\SIrange[range-phrase=--]{5.8}{5.96}{\giga\hertz}.  The LUT is generated using this data
set with $ f_{step} = \SI{10}{\kilo\hertz} $ and $S_G = \{18, 20, 22, 25\}$\footnote{For
the duration of the measurements necessary for LUT construction, the cryogenic switch
was kept pointing towards the noise source.}.  The JPA tuning is then performed as part
of the haloscope experiment following the steps below at each iteration:~:

\begin{enumerate}
    \item Tune the cavity to target frequency $ f_{cav} $.
    \item Let $ f_{sc} $ to be $ f_{cav} $.
    \item Search the LUT for an entry corresponding to the chosen $ f_{sc} $
        and $ G_J = \SI{20}{\decibel} $.  If there is an entry, use the
        corresponding $ (i_b, f_p, P_p) $ to set the working point.  If there
        is no entry, abort the procedure.
    \item The impedance seen by the JPA will be slightly different depending on the
        cryogenic switch position.  Since the LUT is generated with the switch pointing
        at the noise source, the $ G_J $ measured at this step will usually be off by
        about \SIrange[range-phrase=--]{1}{2}{\decibel}.  To compensate, fine-tune $ P_p
        $ until $ G_J $ is \SI[separate-uncertainty=true]{20 \pm 0.4}{\decibel}.  This
        will also correct for the small drifts in the pump signal power.
    \item Measure the noise temperature.
    \item Integrate axion-sensitive spectra.
\end{enumerate}

\begin{figure}[tbh]
    \centering
    \includegraphics[width=0.85\linewidth]{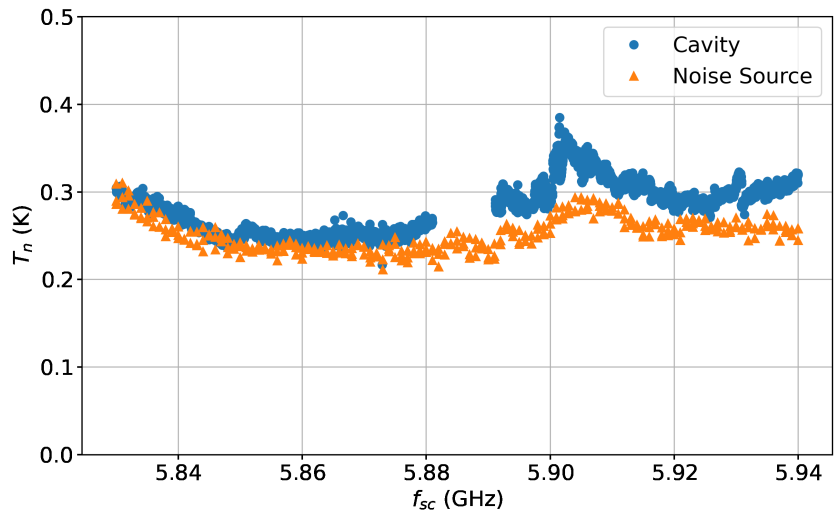}
\caption{$ T_n $ versus operating frequency ($ f_{sc} $) for the detection
chain in the axion experiment.  Blue data points correspond to combined results
of in-situ measurements during data taking runs spanning 6 months.  The orange
data set was measured within a day while the cryogenic switch is pointing at a
\SI{50}{\ohm} resistor at \SI{40}{\milli\kelvin} (noise source).  The major
source of difference between the two measurements is related to the non-ideal
behavior of circulators leading to complex interactions between the cavity
and the JPA.  The noise temperature being higher than the quantum limit of
\SI{140}{\milli\kelvin} is attributed to the accumulated loss and
reflection effects of the components before the JPA with the approximate
contribution of \SIrange{20}{26}{\milli\kelvin} from the later
amplification stages.
}
\label{fig:tncompcavns}
\end{figure}

In addition to in-situ measurements, noise temperatures of selected points from
the LUT were also measured with the cryogenic switch pointing at the noise
source (see Fig.~\ref{fig:tncompcavns}).

\section{Conclusion}
For a JPA based detection chain, the $ T_n $ and $ G_J$ depend on the control
parameters in a non-trivial manner.  
While it is possible to minimize $ T_n $ at a given $ G_J $ by adjusting the
JPA control parameters during the experiment, this is time-consuming and
tedious when the number of working points required is large.
In this work, we proposed a look-up table based method for optimizing the noise
temperature of a JPA at a given gain and operating frequency.
In order to generate a viable LUT, a straightforward characterization protocol
for gain measurements was used. The construction of the LUT relies on
the knowledge that the $ T_n$ is minimized with $ P_p$ which was confirmed by
separate measurements.  The proposed approach was applied successfully in the
context of an axion haloscope experiment around $
\SI{5.9}{\giga\hertz} $.  The LUT generation protocol was fully automated with a 
typical measurement time of about 12 hours for \SI{120}{\mega\hertz} of coverage.
The produced LUT is then viable for the whole duration of a cooldown which
typically lasted more than a month.  During the axion experiment, the JPA
center frequency was tuned from \SIrange{5.83}{5.94}{\giga\hertz} using
\SI{20}{\kilo\hertz} steps while maintaining minimal noise temperature.  The
tuning times were nominally less than \SI{2}{\second} with the limiting factor
being the time cost of measurements during online gain correction.  Moreover,
it was observed that LUTs constructed on separate cooldowns of the cryostat
yielded nearly identical results in noise temperature.  Currently, this method
is employed in all experiments involving a JPA in CAPP\@.

\section{Acknowledgement}
This work is supported in part by the Institute for Basic Science (IBS-R017-D1) and JST
ERATO (Grant No.~JPMJER1601). Arjan F.~van Loo was supported by a JSPS postdoctoral
fellowship.

\end{document}